\documentclass[aps,prl,reprint,groupedaddress,nofootinbib]{revtex4-1}
\usepackage{hyperref}
\usepackage{amsmath}
 \usepackage{multirow}
\usepackage{array}
\newcolumntype{L}[1]{>{\raggedright\let\newline\\\arraybacksslash\hspace{0pt}}m{#1}}
\newcolumntype{C}[1]{>{\centering\let\newline\\\arraybackslash\hspace{0pt}}m{#1}}
\newcolumntype{R}[1]{>{\raggedleft\let\newline\\\arraybackslash\hspace{0pt}}m{#1}}
\usepackage{float}
\usepackage{graphicx}
\usepackage{epsfig}
\usepackage{psfrag}
\usepackage{color}
\usepackage{slashed}

\usepackage{amsfonts}
\usepackage{amssymb}
\usepackage{tikz}
\usepackage{tikz}
\usetikzlibrary{positioning,arrows}
\usetikzlibrary{decorations.pathmorphing}
\usetikzlibrary{decorations.markings}

\newcommand*{\be}{\begin{equation}}
\newcommand*{\ee}{\end{equation}}
\newcommand*{\bea}{\begin{eqnarray}}
\newcommand*{\eea}{\end{eqnarray}}

\newcommand{\comment}[1]{}

\newcommand{\cref}[1]{Chapter~\ref{c.#1}}

\def\beq{\begin{equation}}
\def\eeq{\end{equation}}
\def\bea{\begin{eqnarray}}
\def\eea{\end{eqnarray}}
\def\ba{\begin{array}}
\def\ea{\end{array}}
\def\bi{\begin{itemize}}
\def\ei{\end{itemize}}
\def\be{\begin{enumerate}}
\def\ee{\end{enumerate}}
\def\bc{\begin{center}}
\def\ec{\end{center}}
\def\bt{\begin{table}}
\def\et{\end{table}}
\def\btb{\begin{tabular}}
\def\etb{\end{tabular}}

\def\lsim{\raise0.3ex\hbox{$\;<$\kern-0.75em\raise-1.1ex\hbox{$\sim\;$}}}
\def\gsim{\raise0.3ex\hbox{$\;>$\kern-0.75em\raise-1.1ex\hbox{$\sim\;$}}}
	
\usepackage{cancel}
 \usepackage[normalem]{ulem}
\usepackage{color}
\def\comment#1{\textcolor{blue}{\large(\it{#1})}}

\def\lapp{\mathrel{\rlap{\raise.5ex\hbox{$<$}}
                    {\lower.5ex\hbox{$\sim$}}}}
\def\gapp{\mathrel{\rlap{\raise.5ex\hbox{$>$}}
                    {\lower.5ex\hbox{$\sim$}}}}

\begin{document}

\title{Exploring the sensitivity of hadron colliders to non-universality in heavy neutral currents}

\author{F. A. Conventi$^{1,2}$}
\email{francesco.conventi@cern.ch}
\author{G. D'Ambrosio$^{1}$}
\email{gdambros@na.infn.it}
\author{A.M.  Iyer$^{3}$}
\email{a.iyer@ipnl.in2p3.fr,abhishekiyer1@gmail.com}
\thanks{Present Address: Department of Physics, IIT Delhi, Hauz Khas New Delhi-110016, India}
\author{E. Rossi$^{1,4}$}
\email{Elvira.Rossi@cern.ch}

\affiliation{$^1$INFN-Sezione di Napoli, Via Cintia, 80126 Napoli, Italy;\\
$^2$Universit\`a degli Studi di Napoli Parthenope, Napoli, Italy;\\
$^3$Univ. Lyon, Universite Claude Bernard Lyon 1, CNRS/IN2P3, UMR5822 IP2I,
F-69622, Villeurbanne, France;\\
$^4$ Universit\`a degli Studi di Napoli ’Federico II’, Dipartimento di Fisica “Ettore Pancini”, Via Cintia, 80126 Napoli, Italy.}


\begin{abstract}
We present sensitivity projections for discovering a heavy resonance decaying to electron and muon pairs and for probing the charged lepton non-universality in such decays at the  HL-LHC and FCC-hh. The analysis takes into account the expected differences in the reconstruction efficiencies and the dilepton mass resolutions for dielectron and dimuon final states. We demonstrate how the analyses at HL-LHC  naturally paves the way for a FCC-hh machine thereby underlining its importance.

\end{abstract}

\maketitle

\noindent 
The Standard Model (SM) of particle physics has withstood the test of experimental validation to a significant extent.
In particular, the electroweak sector is characterized by a well defined pattern of couplings which manifests in terms of accurate predictions for several processes. Any departure from this paradigm implies the presence of New Physics (NP) effects.
One of the most interesting observations in this direction corresponds to the observation of flavour non-universality in terms of the theoretically clean ratios:  $R_K$ and $R_{K^*}$. 
Recent results obtained by the LHCb Collaboration are compatible with the standard model at the level of 2.5 standard deviations \cite{Aaij:2019wad,Aaij:2017vbb}, still leaving room for studies on flavour non-universality.
Anyway, independently of these anomalies, it is instructive to investigate the potential of the direct searches in measuring deviations from universality. 

In a direct search, non-universality between a set of final states would manifest in the form of correspondingly different yields in the detector. 
In this paper we present sensitivity projections for testing charged lepton flavor non-universality in dilepton decays of a new heavy boson that should be discovered at current and future $pp$ colliders.
The analysis uses a simple test statistic to estimate the significance of these departure from the universality case.

Since the leptons are very clean objects in a detector, the developed strategy will be used to study flavour non-universality using a simplified model with an additional heavy state. Without loss of generality, we consider a heavy vector boson ($Z'$) decaying into a pair of leptons. 
These states are a characteristic of several models beyond the SM: for instance scenarios with additional $U(1)$ \cite{Donini:1997yu}, extra-dimensional frameworks with bulk
gauge fields \cite{Gherghetta:2000qt} constitute some of the most obvious extensions. The scenarios with an additional heavy vector were also found to be useful in the context of  flavour physics \cite{Gauld:2013qba,Glashow:2014iga,Bhattacharya:2014wla,Crivellin:2015mga,Crivellin:2015lwa,Sierra:2015fma,Crivellin:2015era,Celis:2015ara,Belanger:2015nma,Gripaios:2015gra,Allanach:2015gkd,Fuyuto:2015gmk,Chiang:2016qov,Boucenna:2016wpr,Boucenna:2016qad,Celis:2016ayl,Altmannshofer:2016jzy,Crivellin:2016ejn,GarciaGarcia:2016nvr,Becirevic:2016zri,Bhattacharya:2016mcc,Bhatia:2017tgo,Cline:2017lvv}.
The analysis in this paper, however, can be trivially extended to the $s$-channel decay of any heavy resonance like gravitons, heavy scalars \textit{etc.}
The production mechanism is irrelevant for our analysis, therefore, without loss of generality, we assume the $Z'$ being predominantly produced by light quarks. From a model point of view, denoting the coupling of the leptons to the vector boson as $g_l$, the goal of this paper can be restated in terms of extracting the sensitivity of the direct searches to explore the difference $g_e-g_\mu$ and its deviations from 0. Similar analyses exist for the Z-boson couplings from LEP \cite{ALEPH:2005ab}. \footnote{The analysis presented in this paper can be easily applied to test universality of couplings for the SM Z boson as well.}

The paper is organized as follows: we begin with the traditional bump hunt searches of heavy neutral resonances decaying into a di-lepton (di-muon and di-electron) final state. In this section we point out the role of the different reconstruction resolutions between different flavour leptons in the eventual computation of the discovery significance. This is then followed by the description of the analysis and the estimation of the sensitivity to non-universality of the HL-LHC collider. We note that the study at HL-LHC naturally paves the way for an FCC-hh machine which is characterized by significantly enhanced sensitivities to even smaller deviations from universality. We conclude the paper with the prospects of including tau as a part of future analysis to complete the picture.
%
%
\section{Bump Hunt Searches}
The search for a heavy neutral resonance decaying into a di-lepton final state is one of the most prominent channels being probed at LHC and there exist relatively strong bounds on $\sigma\times \mathcal{B}_{ll}$ \cite{Sirunyan:2018exx,Aad:2019fac}.
A standard search strategy focuses on the possibility for observing an excess of events over the Standard Model (SM) prediction, where the SM background is mainly due to the universal coupling of the $\gamma^*/Z$ to leptons.

In this analysis, we consider the  production of a heavy $Z'$ decaying into muons and electrons. 
 For the purpose of simulation, we use the Lagrangian of the Sequential Standard Model (SSM).
  Using the model file from {\tt{FEYNRULES}} \cite{Alloul:2013bka}, the matrix element for the process is produced using {\tt{MADGRAPH}} \cite{Alwall:2014hca} at a centre of mass energy of 14 TeV. Showering and hadronization are described using {\tt{PYTHIA 8}} \cite{Sjostrand:2007gs}. CMS cards of {\tt{DELPHES 3.4}} \cite{deFavereau:2013fsa} is used for detector simulation at the LHC.
 The efficiencies estimated from the simulation are then used for different values of $\sigma\mathcal{B}$.

  

\textbf{Event selection:} In order to identify the leptons from the $Z'$, the following selection criteria have been applied:
\begin{itemize}
	\item two isolated leptons (electrons or muons) with a $p_T\ge50~GeV$ and $|\eta| <2.5$;
	\item $\slashed E_T < 10 GeV$.
\end{itemize}
The main source of background is represented by the $pp\rightarrow Z/\gamma^*\rightarrow ll$ where $l=e,\mu$. 

\begin{figure}
	\begin{center}
		\includegraphics[width=6cm]{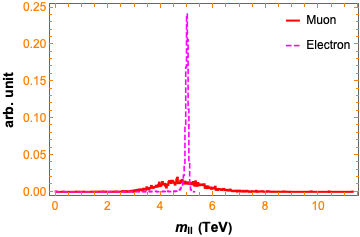}
	\end{center}
	\caption{Mass reconstruction resolution of the di-electron  (in pink-dashed) and the di-muon (in red-solid) pairs for $M_{Z'}=5$ TeV. 	}
	\label{fig:narrowvsbroad}
\end{figure}

Independently of the relative sizes of the coupling with the vector boson (SM or beyond), the leptons are characterized by different detector acceptances and mass reconstruction resolution. The acceptance efficiency ($\epsilon$) is mass dependent: For instance for $m_{Z'}=3$ TeV, we estimate $\epsilon_e=0.46$ and  $\epsilon_\mu=0.61$. While for 5 TeV the corresponding values for electrons and muons are 0.48 and 0.35 respectively. The mass reconstruction resolution for $m_{ll}> 1$~TeV is much better for the di-electron final state.
The mass reconstruction resolution for di-leptons is shown in Fig. \ref{fig:narrowvsbroad} for a 5 TeV narrow resonance with a generated mass width $\Gamma$ of $50$ GeV. 
The different mass reconstruction resolution can be attributed to fact that the momentum of the electrons and muons are measured differently: the former due to deposition in the E-cal and the latter due to the bending in the tracker.\footnote{Under the assumption of enough statistics (not necessarily equal) for either lepton, the asymmetry in the reconstruction between the electrons and muons progressively increases with the resonance mass. The smearing increases with the $p_T$ of the di-muons.}

%
%
%
%

To calculate the expected significance for $Z'\rightarrow ee$ and $Z'\rightarrow \mu\mu$ at LHC, we use
	a binned likelihood fit $L(\mu_e,\mu_\mu)$. 

In the case where background is well known 
we can evaluate the expected significance as the probability of background only hypothesis ($\mu_e=\mu_\mu = 0$) using the two dimensional profiled likelihood ratio test \cite{Cowan:2010js}:
\begin{equation}
q_0=-2\log\left[\frac{ L( 0, 0)}{L( \hat \mu_e,\hat \mu_\mu)}\right]
\label{eq:discovery}
\end{equation}
where $\hat \mu$ is the best value of $\mu$ estimated by fitting to the data for both the electron and the muon. The signal discovery significance Z can be evaluated as:
\begin{equation}
Z_{tot}=\sqrt{q_0}.
\end{equation}
and for sufficiently large background we can use the asymptotic formula:
\begin{equation}
Z_{tot}=\sqrt{q_0}=\sqrt{\sum\limits_{i=1;j=e,\mu}^{N_{e,\mu}} \left(2(s^j_i+b^j_i)\log\left[1+\frac{s^j_i}{b^j_i}\right]-2s^j_i\right)}
\label{eq:sensitivity}
\end{equation}
where the sum runs over the bins, $s^j_i$ and $b^j_i$ are the expected numbers for signal and background events in the $i^{th}$ bin for $j=e,\mu$. Note that the total number of bins $N_{e,\mu}$ are in general different for the electron and the muon. It is important to stress that Eq. \ref{eq:sensitivity} just gives the local significance. We account for the look elsewhere effect which leads to the modification of local p-value corresponding to a given $Z_{tot}$ \cite{LEE:2010epj}. 

%
%
\begin{figure}
	\includegraphics[width=6cm]{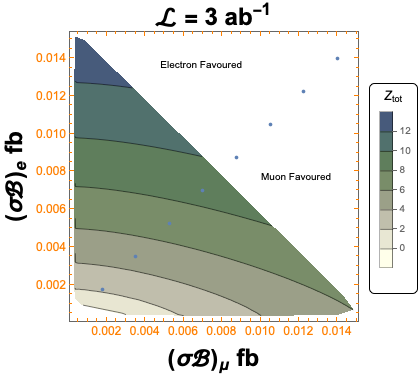}\\
	
	\caption{ 
			Contours in the total di-lepton significance ($Z_{tot}$) for $m_{Z'}=5$ TeV decaying into electrons $(\sigma \mathcal{B})_e $ and muons $(\sigma \mathcal{B})_\mu $. The diagonal dotted line corresponds to the lepton flavour universality case $\left( (\sigma\mathcal{B})_e =(\sigma\mathcal{B})_\mu \right )$. 
			The asymmetry in the expected significance is due to different mass reconstruction resolution (Fig.\ref{fig:narrowvsbroad}).}
	
	\label{fig:BeBmu}
\end{figure}
Fig. \ref{fig:BeBmu} gives contours in the total di-lepton significance as a function of cross-section times the branching fractions of the $Z'$ decaying into electrons $(\sigma \mathcal{B})_e$ and muons $(\sigma \mathcal{B})_\mu $. 
The diagonal dotted line corresponds to the lepton flavour universality case $\left( (\sigma\mathcal{B})_e =(\sigma\mathcal{B})_\mu \right )$. 
The points are scanned such that $(\sigma\mathcal{B})_e+(\sigma\mathcal{B})_\mu\leq (\sigma\mathcal{B})_{max}$, with the outer edge corresponding to $(\sigma\mathcal{B})_{max}$. For the LHC, we choose $(\sigma\mathcal{B})_{max}$ as the upper bound obtained on $(\sigma\mathcal{B})_{tot}$ from direct searches in di-lepton final state \cite{Sirunyan:2018exx,Aad:2019fac}. Lines parallel to the outer edge represent contours of some constant $(\sigma\mathcal{B})_{tot}<(\sigma\mathcal{B})_{max}$, decreasing progressively as one moves inwards.  For any given $(\sigma\mathcal{B})_{tot}$, the scan over $(\sigma\mathcal{B})_{e}$ and $(\sigma\mathcal{B})_{\mu}$ is done such that $(\sigma\mathcal{B})_{e}+(\sigma\mathcal{B})_{\mu}=(\sigma\mathcal{B})_{tot}$
The asymmetric behaviour of the contour plot is due to the different mass resolutions shown in Fig. \ref{fig:narrowvsbroad}.
Thus, a larger coupling to the electrons leads to a larger evaluated value for the total signal sensitivity. 
These considerations lead to the following questions:
\begin{enumerate}
    \item does the absence of a signal imply no NP or a larger coupling to the muons?
    \item what are the prospects for unearthing non-universality at the HL-LHC and future colliders?
\end{enumerate}
%
%
%
%
\section{Non-universality test}
\begin{figure}
	\centering	
	\begin{tabular}{cc}
		\includegraphics[width=4.2cm]{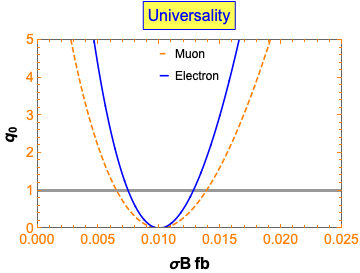}&\includegraphics[width=4.2cm]{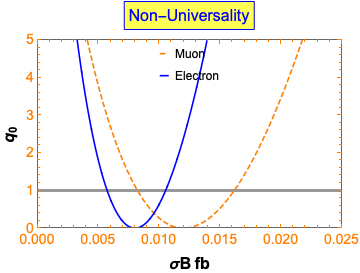}\\
	\end{tabular}
	\caption{Distribution of test statistic under the assumption of universal ($(\sigma\mathcal{B})_e=(\sigma\mathcal{B})_\mu$) (left plot) and non-universal ($(\sigma\mathcal{B})_e<(\sigma\mathcal{B})_\mu$) (right plot) couplings. The different widths are a consequence of different mass reconstruction resolution for the electrons (blue-thick) and muons (orange-dashed). }
	\protect\label{fig:LLR}
\end{figure}

In real life experiments, the statistic $q_0$ in Eq. \ref{eq:discovery} is minimized at the best fit value of $\sigma\mathcal{B}$ for the leptons.
Fig. \ref{fig:LLR} shows the distributions of the test statistic $q_0$ under two different assumptions: the left plot corresponds to the universal coupling case where $(\sigma\mathcal{B})_e=(\sigma\mathcal{B})_\mu$ while the right plot illustrates the $(\sigma\mathcal{B})_e<(\sigma\mathcal{B})_\mu$ case and hence non-universality. 
The different widths of the parabola reflect the differences in the mass reconstruction resolutions between the leptons. The black line represents the 1 $\sigma$ measurement uncertainty.
%
%
%
%
%
The departure from the universality hypothesis can be quantified by  the following asymmetry variable:
\begin{equation}
\hat A =\frac{ (\sigma \mathcal{B})_{\mu}-(\sigma \mathcal{B})_{e}}{ (\sigma \mathcal{B})_{\mu}+(\sigma \mathcal{B})_{e}} \in [-1,1].
\label{eq:asymm}
\end{equation}
The two extremities $\hat A=-1$ and $\hat A=1$ correspond to a very large signal in the electron channel ($\sigma \mathcal{B})_{e}\gg(\sigma \mathcal{B})_{\mu}$ and muon channel ($\sigma \mathcal{B})_{\mu}\gg(\sigma \mathcal{B})_{e}$ respectively. In general, $\hat A$ divides the phase space into two specific regions: $\hat A>0$ corresponds to the case where couplings to muons is larger and is called the \textit{Pro-muon} region; $\hat A<0$ corresponds to the case where couplings to electrons is larger and is called the \textit{Pro-electron} region.
Thus, a measurement corresponding to $\hat{A}\neq0$ could be a hint of non-universality.
An estimate for the significance in the measurement of $\hat{A}$ must also account for the individual uncertainties in the extraction of $\sigma\mathcal{B}_{e,\mu}$ which correspond to the widths in Fig. \ref{fig:LLR}. 
	
	The significance in the measurement of $\hat{A}$ can be quantified by using a two dimensional profiled likelihood ratio test similar to Eq. \ref{eq:discovery} and defined as
	\begin{equation}
q=-2\log\left[\frac{ L( \hat{A} = 0)}{L( \hat{A} )}\right]
\label{eq:Var}
\end{equation}
treating $(\sigma\mathcal{B})$ as a nuisance parameter.
The measured values of $(\sigma\mathcal{B})$ for the electrons and muons are related to $(\sigma\mathcal{B})_{tot}$ as:
	$(\sigma\mathcal{B})_{e,\mu}=\hat \mu^m_{e,\mu}(\sigma\mathcal{B})_{tot}$.

\begin{figure}
	\begin{center}
		\begin{tabular}{cc}
			\includegraphics[width=4.2cm]{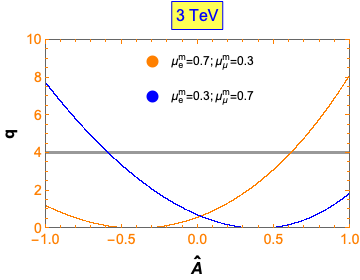} &	\includegraphics[width=4.2cm]{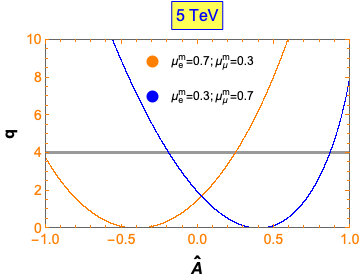}\\	\includegraphics[width=4.2cm]{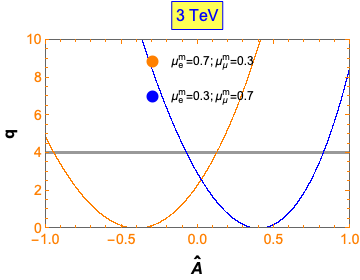} &	\includegraphics[width=4.2cm]{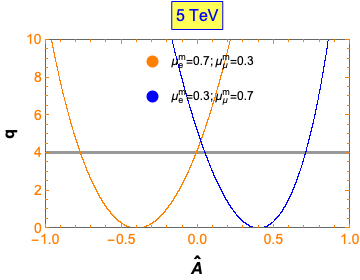}
		\end{tabular}
	\end{center}
	\caption{Test statistic $q$  in Eq. \ref{eq:Var} with a benchmark of $\sigma\mathcal{B}_{Tot}=0.01$ fb (top row) and $0.025$ fb ( bottom row). Left (right) column corresponds to $M_{Z'}=3(5)$ TeV. The orange and blue curves correspond to two different hypothesis for $\mu^m_{e}$ and $\mu^m_{\mu}$ (See text for details).}
	\protect\label{fig:teststat}
\end{figure}

\begin{figure}[htb!]
\centering		\begin{tabular}{c}
			\includegraphics[width=8.2cm]{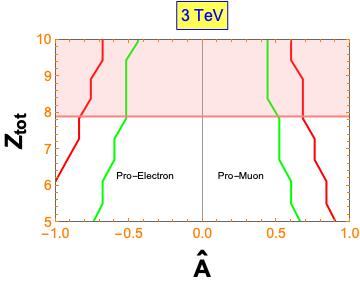}\\ \includegraphics[width=8.2cm]{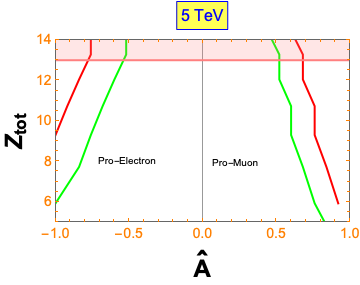}\\
		\end{tabular}
	\caption{2$\sigma$ (Green lines) and 3$\sigma$ (Red solid lines) asymmetry sensitivity plot for the full LHC dataset at $\mathcal{L}=3$ab$^{-1}$. The pink-shaded region is the upper bound on $(\sigma\mathcal{B})_{tot}$ from direct searches at LHC. 
	}
	\protect\label{fig:LHCexclusion}
\end{figure}

Fig. \ref{fig:teststat} shows the typical behaviour of $q$ for two different values of non-universality: $\mu^m_{e}=0.7;\mu^m_{\mu}=0.3$ (orange) and $\mu^m_{e}=0.3;\mu^m_{\mu}=0.7$ (blue).  We use two different  benchmark values of the total cross-section:  $(\sigma\mathcal{B})_{Tot}=0.01$ fb (top row) and $(\sigma\mathcal{B})_{Tot}=0.025$ fb (bottom row) for $M_{Z'}=3,5$ TeV. The plots quantify the departure of the universality hypothesis ($\hat A=0$ or $\hat\mu_e=\hat\mu_\mu=0.5$). The solid black line corresponds to the $2\sigma$ intercept. Scanning for different values of $(\sigma \mathcal{B})_{Tot}$, we obtain the asymmetry-sensitivity plots  shown in  Fig. \ref{fig:LHCexclusion}. 
 Moving along either curve, from the bottom to the top, corresponds to increasing values of $(\sigma \mathcal{B})_{Tot}$ and, hence, $Z_{tot}$. 
 With respect to bounds from direct searches \cite{Aad:2019fac}, we must note that they are obtained under a lepton flavour universality assumption. Estimation of a bound under the non-universality hypothesis will require a recast of the entire analysis and is out of scope of the paper. However, we naively calculate the $Z_{tot}$ from the upper bound on $(\sigma\mathcal{B})_{tot}$ represented by the lower boundary of the orange-shaded region in Fig. \ref{fig:LHCexclusion}.

\begin{figure}
	\begin{center}
		\begin{tabular}{c}
			\includegraphics[width=6.0cm]{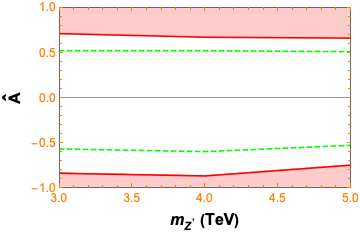}
		\end{tabular}
	\end{center}
	\caption{Summary of exclusion in $\hat {A}$ for different masses corresponding to the rescaled LHC bounds for the electron channel. The red(green-dashed) lines represent 3(2)$\sigma$ exclusion for the full LHC dataset at $\mathcal{L}=3$ab$^{-1}$.}
	\protect\label{fig:LHCsummary}
\end{figure}

In an ideal scenario, for any given Z' mass we can expect to be sensitive to tiny deviations from universality i.e. $\hat A \rightarrow 0$ as $(\sigma\mathcal{B})$ becomes very large or $Z_{tot}$ increase.
However HL-LHC sensitivity is limited by existing bounds on $(\sigma\mathcal{B})_{tot}$ as well as a finite total integrated luminosity. The ruled out region for 3 and 5 TeV masses are illustrated by the pink band in Fig.\ref{fig:LHCexclusion}.
Taking this into account, Fig.\ref{fig:LHCsummary} illustrates the expected limits on $|\hat A|$ as a function of Z' mass for the full LHC dataset at $\mathcal{L}=3$ab$^{-1}$.  The flat behaviour is due to the fact that the bounds on direct searches becomes progressively stronger in going from 1 to 5 TeV.
Having laid out our strategy for extracting non-universality, we find it relevant to draw the attention to Fig. 1 of \cite{Greljo:2017vvb} which uses differential LFU (Lepton flavour universality) ratios to extract non-universality. Similar ratios are also employed by experiments \cite{Aad:2020ayz} and a combination with the proposed analysis in this paper could possibly reveal more information. 
%

%
%

The current non universality tests at LHC, while being powerful are limited on the following accounts:  1) reduced sensitivity to heavier masses 2) Reduced sensitivity to minor deviations from universality. These considerations naturally lead to evaluate and study the possible improvements with future colliders as it will be discussed below.
%
%

%
%
%
\section{Future Colliders}
The advent of the FCChh is expected to provide continuity from the tail end of the sensitivity of the HL-LHC. 
The higher energy and integrated luminosity of FCChh will allow one to extend the discovery reach toward both higher masses and smaller couplings, and to increase the sensitivity for charged lepton flavour non-universality.
This makes a future collider all the more relevant not only for explorations deeper in the UV regions of phase space but also provides enhanced sensitivity to minor deviations from non-universality.

\begin{figure}[htb!]
	\begin{center}
		\begin{tabular}{cc}
			\includegraphics[width=4.2cm]{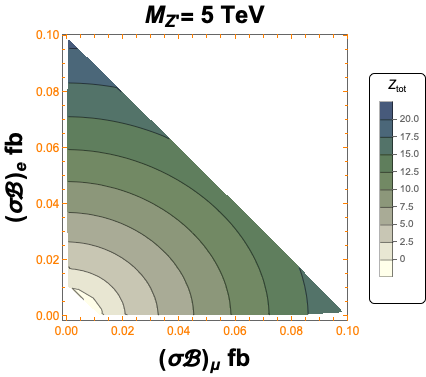}&\includegraphics[width=4.2cm]{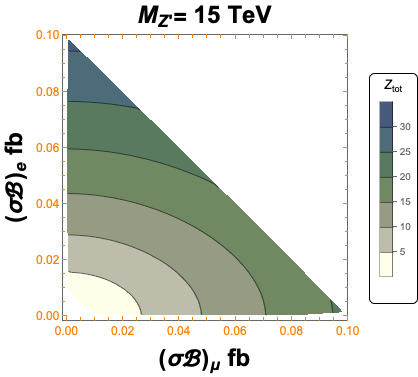}\\
		\end{tabular}
	\end{center}
	\caption{Contours of total signal significance as a function of branching fraction into the leptons for 5 TeV (left) and 15 TeV (right) computed at $\mathcal{L}=10$ ab$^{-1}$ for the FCC. }
	\protect\label{fig:reachFCC}
\end{figure}

We first begin with the discovery prospect of such states at the FCC. For the purpose of FCC studies, we use the FCC-hh card reported in this reference \cite{deFavereau:2013fsa}.
One notable difference between the HL-LHC detectors and those contemplated for FCC-hh is that the latter are expected to have relatively similar reconstruction resolutions for electron and muon momenta. This is particularly true for lower masses as compared to the heavier masses as shown in Fig. \ref{fig:reachFCC} where the effect of the Z' mass resolution on the expected signal sensitivity for the FCC have been reported. 
The philosophy is exactly similar to the corresponding plot for the LHC in Fig. \ref{fig:BeBmu}. All the point on the edge of the plot satisfy $ (\sigma\mathcal{B})_e +(\sigma\mathcal{B})_\mu= 0.1$ fb.
The behavior of increasing asymmetry between the leptons is similar to that of the LHC albeit at much higher masses: at the LHC one expects a symmetric reconstruction at around the scale of the $Z$ boson mass with the asymmetry increasing progressively. 

\begin{figure}[htb!]
	\begin{center}
		\begin{tabular}{c}
			\includegraphics[width=6.0cm]{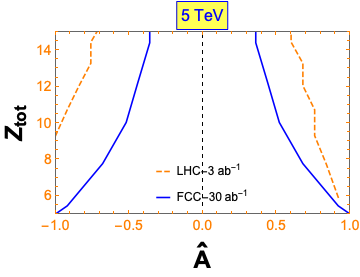}
		\end{tabular}
	\end{center}
	\caption{3$\sigma$ asymmetry sensitivity plot with the respect to expected discovery significance in the electron channel for FCC (blue-solid) at a luminosity of 30 ab$^{-1}$ and for LHC (orange-dashed)) at a luminosity of 3 ab$^{-1}$ for a $Z'$ mass of 5 TeV.}
	\protect\label{fig:FCCLHCcomparison}
\end{figure}

Our procedure to evaluate the signal discovery significance for FCC will be exactly similar to the one shown for HL-LHC. For the purpose of continuity we begin with $M_{Z'}=5$ TeV and compare the FCC results with that at HL-LHC as shown in Fig.\ref{fig:FCCLHCcomparison}. 

\begin{figure}
	\begin{center}
		\begin{tabular}{c}
			\includegraphics[width=8.2cm]{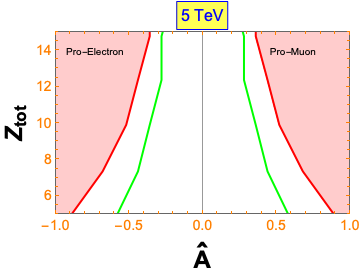}\\	\includegraphics[width=8.2cm]{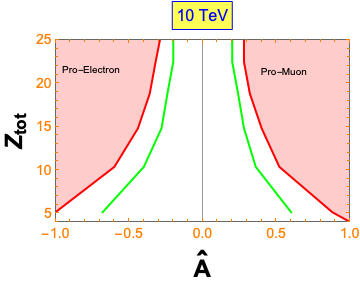}\\
		\end{tabular}
	\end{center}
	\caption{3 $\sigma$ (red shadowed region) and 2 $\sigma$ (green lines)  asymmetry sensitivity plot for the FCC with the respect to expected discovery significance in the electron channel for a $Z'$ mass of 5 and 10 TeV with a luminosity of 30 ab$^{-1}$. The red shadowed region illustrates regions of  non-universality that can be excluded at $3\sigma$ at this luminosity. }
	\protect\label{fig:FCCexclusion}
\end{figure}

The results are compared at the end of the expected run of the corresponding machines: 3 ab$^{-1}$ for HL-LHC and 30 ab$^{-1}$ for FCC. All the lines represent 3 $\sigma$ sensitivity to non-universality.
The FCC curve is characterized by two distinct features: A) more symmetric sensitivity on either side of $\hat A$ owing to symmetric reconstruction for 5 TeV $Z'$ mass and B) higher sensitivity to minor deviations from non-universality corresponding to the regions around $\hat A=0$.
%
%
FCC results for a $Z'$ mass of 5 and 10 TeV are shown in Fig. \ref{fig:FCCexclusion}.

As noted before, the strength of the FCC at 30 ab$^{-1}$ is demonstrated by its ability to probe regions very close to $\hat{A}=0$.
%
%
Table \ref{tab:tab1} summarize the A bounds for a Z' of 5 TeV and 10 TeV for Zee equal to 10 and 15. for HL-LHC and FCC.

\begin{table}[h]
	\centering
	\begin{tabular}{|c|c|c|c|}
		\hline
		$m_{Z'}$ & $Z_{tot}$ & HL-LHC & FCC\\
		\hline
		$5$ TeV & 10              & $(-0.95, 0.76)$ & $(-0.53, 0.52)$ \\
		& 15              & $--$ & $(-0.36, 0.35)$ \\
		
		\hline
		\hline
		
		$10$ TeV & 10              &-- & $(-0.63, 0.55)$\\
		
		& 15              &-- & $(-0.42, 0.38)$\\
		\hline
	\end{tabular}
	\caption{3$\sigma$ $\hat{A}$ bounds for a $5$ and $10$ TeV $Z'$ at $Z_{ee}=10,15$ level for HL-LHC and FCC. 
		We do not quote the sensitivity for 10 TeV at HL-LHC as it is  out of reach for practical values of $\sigma\mathcal{B}$}. \label{tab:tab1}
\end{table}

\section{Conclusions}
In this work, using a simple test statistic, we present sensitivity projections for testing charged lepton flavour universality in dilepton decays of a neutral heavy boson, should it be discovered at HL-LHC or FCC-hh. While being powerful HL-LHC limits show reduced sensitivity to heavier Z' masses and minor deviations from leptons universality.
This motivates a detailed analysis in the future FCC-hh machines where the differences between the electrons and the muons are ironed out for relatively lighter masses. Furthermore, this machine is sensitive to minor deviations from non-universality. This work also offers a nice complementarity between the observations in flavour factories and direct searches. This strategy can also be extended to tau final states which are mainly identified by their hadronic decays. Using the techniques introduced in this paper and adapting improved identification criteria for the tau, will enable us to get a complete picture of (non-)universality in the neutral current sector. 
\vspace{.3cm}
\section{Acknowledgements}
We are grateful to M. Mangano for his continuous suggestions throughout the course of the project.
F.C. and A.I would wish to thank  Antonio Giannini  for his help with the computation of signal sensitivities in the initial stages of the project.
G.D and A.I. wish to thank useful discussions with Alberto Orso Maria Iorio.
AI wishes to thank Michael Winn for useful observations during the GdR-InF 2019 meeting.
We wish to thank Sabyasachi Chakraborty, Seema Sharma and Tuhin Roy for a careful reading of the manuscript and several useful comments.  A.I would like to thank CEFIPRA under the project “Composite Models at the
Interface of Theory and Phenomenology” (Project No.
5904-C).
G.D. was supported in part by MIUR under Project No. 2015P5SBHT and by the INFN research initiative ENP. G.D. thanks “Satish Dhawan Visiting Chair Professorship” at the Indian Institute of Science.




\bibliography{biblio1}
\end{document}